\documentstyle[12pt,epsfig]{article}
\newcounter{subequation}[equation]

\makeatletter 
 
\def\thesubequation{\theequation\@alph\c@subequation}
\def\@subeqnnum{{\rm (\thesubequation)}}
\def\slabel#1{\@bsphack\if@filesw {\let\thepage\relax
   \xdef\@gtempa{\write\@auxout{\string
      \newlabel{#1}{{\thesubequation}{\thepage}}}}}\@gtempa
   \if@nobreak \ifvmode\nobreak\fi\fi\fi\@esphack}
\def\subeqnarray{\stepcounter{equation}
\let\@currentlabel=\theequation\global\c@subequation\@ne
\global\@eqnswtrue
\global\@eqcnt\z@\tabskip\@centering\let\\=\@subeqncr
$$\halign to \displaywidth\bgroup\@eqnsel\hskip\@centering
  $\displaystyle\tabskip\z@{##}$&\global\@eqcnt\@ne
  \hskip 2\arraycolsep \hfil${##}$\hfil
  &\global\@eqcnt\tw@ \hskip 2\arraycolsep
  $\displaystyle\tabskip\z@{##}$\hfil
   \tabskip\@centering&\llap{##}\tabskip\z@\cr}
\def\endsubeqnarray{\@@subeqncr\egroup
                     $$\global\@ignoretrue}
\def\@subeqncr{{\ifnum0=`}\fi\@ifstar{\global\@eqpen\@M
    \@ysubeqncr}{\global\@eqpen\interdisplaylinepenalty \@ysubeqncr}}
\def\@ysubeqncr{\@ifnextchar [{\@xsubeqncr}{\@xsubeqncr[\z@]}}
\def\@xsubeqncr[#1]{\ifnum0=`{\fi}\@@subeqncr
   \noalign{\penalty\@eqpen\vskip\jot\vskip #1\relax}}
\def\@@subeqncr{\let\@tempa\relax
    \ifcase\@eqcnt \def\@tempa{& & &}\or \def\@tempa{& &}
      \else \def\@tempa{&}\fi
     \@tempa \if@eqnsw\@subeqnnum\refstepcounter{subequation}\fi
     \global\@eqnswtrue\global\@eqcnt\z@\cr}
\let\@ssubeqncr=\@subeqncr
\@namedef{subeqnarray*}{\def\@subeqncr{\nonumber\@ssubeqncr}\subeqnarray}
\@namedef{endsubeqnarray*}{\global\advance\c@equation\m@ne%
                           \nonumber\endsubeqnarray}

\makeatletter
\@addtoreset{equation}{section}
\makeatother
\renewcommand{\theequation}{\thesection.\arabic{equation}}

\def\dalemb#1#2{{\vbox{\hrule height .#2pt
        \hbox{\vrule width.#2pt height#1pt \kern#1pt
                \vrule width.#2pt}
        \hrule height.#2pt}}}

    \let\e=\epsilon
  \let\q=\theta  
  \let\n=\nu

 \def\bd{\begin{document}} \def\ed{\end{document}}
\def\ds{\documentstyle} \let\fr=\frac \let\bl=\bigl \let\br=\bigr
\let\Br=\Bigr \let\Bl=\Bigl 
\let\bm=\bibitem
\let\na=\nabla
\let\pa=\partial \let\ov=\overline
\def\ie{{\it i.e.\ }} 
\newcommand{\be}{\begin{equation}} 
\newcommand{\ee}{\end{equation}} 
\def\ba{\begin{array}}
\def\ea{\end{array}}
\def\ft#1#2{{\textstyle{{\scriptstyle #1}\over {\scriptstyle #2}}}}
\def\fft#1#2{{#1 \over #2}}
\def\del{\partial}
\def\sst#1{{\scriptscriptstyle #1}}
\def\oneone{\rlap 1\mkern4mu{\rm l}}
\def\e7{E_{7(+7)}}
\def\td{\tilde}
\def\wtd{\widetilde}
\def\im{{\rm i}}
\def\bog{Bogomol'nyi\ }
\def\q{{\tilde q}}
\def\hast{{\hat\ast}}
\def\0{{\sst{(0)}}}
\def\1{{\sst{(1)}}}
\def\2{{\sst{(2)}}}
\def\3{{\sst{(3)}}}
\def\4{{\sst{(4)}}}
\def\5{{\sst{(5)}}}
\def\6{{\sst{(6)}}}
\def\7{{\sst{(7)}}}
\def\8{{\sst{(8)}}}
\def\n{{\sst{(n)}}}
\def\oo{{\"o}}
\def\hA{\hat{\cal A}}
\def\ns{{\sst {\rm NS}}}
\def\rr{{\sst {\rm RR}}}
\def\tH{{\widetilde H}}
\def\tB{{\widetilde B}}
\def\cA{{\cal A}}
\def\cF{{\cal F}}
\def\tF{{\wtd F}}
\def\Z{\rlap{\sf Z}\mkern3mu{\sf Z}}
\def\ep{{\epsilon}}
\def\IIA{{\rm IIA}}
\def\IIB{{\rm IIB}}
\def\ads{{\rm AdS}}
\def\R{\rlap{\rm I}\mkern3mu{\rm R}}
\def\mapright#1{\smash{\mathop{-\!\!\!-\!\!\!-\!\!\!-\!\!\!-\!\!\!
             \longrightarrow}\limits^{#1}}}

\def\Ei{{\hbox{Ei}}}
\def\Ci{{\hbox{Ci}}}
\def\Si{{\hbox{Si}}}

\newcommand{\ho}[1]{$\, ^{#1}$}
\newcommand{\hoch}[1]{$\, ^{#1}$}
\newcommand{\bea}{\begin{eqnarray}} 
\newcommand{\eea}{\end{eqnarray}} 
\newcommand{\ra}{\rightarrow}
\newcommand{\lra}{\longrightarrow}
\newcommand{\Lra}{\Leftrightarrow}
\newcommand{\aap}{\alpha^\prime}
\newcommand{\bp}{\tilde \beta^\prime}
\newcommand{\tr}{{\rm tr} }
\newcommand{\Tr}{{\rm Tr} } 
\newcommand{\NP}{Nucl. Phys. }
\newcommand{\tamphys}{\it Center for Theoretical Physics,
Texas A\&M University, College Station, TX 77843}

\newcommand{\upenn}{\it Department of Physics and Astronomy,\\ University
of Pennsylvania, Philadelphia, PA 19104}

\newcommand{\brussels}{\it Physique Th\'eorique et Math\'ematique, 
Universit\'e Libre de Bruxelles,\\ Campus Plaine C.P. 231, B-1050
Bruxelles, Belgium} 

\newcommand{\auth}{J. F. V\'azquez-Poritz}

\begin{document}
\begin{flushright}
ULB-TH/01-31\\
UPR/ 961-T\\
October  2001\\
\hfill{\bf hep-th/0110085}\\
\end{flushright}


\begin{center}

{\large {\bf Investigating Brane Resolution With Perturbative Dynamics}}

\vspace{20pt}

\auth

\vspace{10pt}
\brussels\\
\vspace{10pt}
\upenn

\vspace{30pt}

\underline{ABSTRACT}
\end{center}

We investigate the transition between singular and non-singular geometries
from the vantage point of perturbative field dynamics. In particular, we
obtain the closed-form absorption probability for minimally-coupled
massless scalars propagating in the background of a heterotic 5-brane on a
Taub-NUT instanton. This is an exact calculation for arbitrary incident
frequencies. For the singular geometry, the absorption probability vanishes
when the frequency is below a certain threshold, and for the non-singular
case it vanishes for all frequencies. We discuss the connection between this
phenomenon and the behavior of geodesics in this background. We also obtain
exact quasinormal modes.

{\vfill\leftline{}\vfill
\vskip 10pt \footnoterule {\footnotesize \hoch{1} This work is supported
in part by the Francqui Foundation (Belgium), the Actions de Recherche
Concert{\'e}es of the Direction de la Recherche Scientifique - Communaut\'e
Francaise de Belgique, IISN-Belgium (convention 4.4505.86), and DOE grant
DOE-FG02-95ER40893
 
\vskip  -12pt} \vskip   14pt
}

\pagebreak
\setcounter{page}{1}


\section{Introduction}

In certain cases, deformations of the standard brane solutions can have the 
effect of "resolving" singularities. Such deformations are the result of an
additional flux on a Ricci-flat space transverse to the brane. Since this
resolution can break additional supersymmetry, such non-singular solutions
may serve as viable gravity duals of strongly-coupled Yang-Mills field
theories with less than maximal supersymmetry
\cite{kleb,strass,grana,gub,clp,cglp}. 

In this paper, we look at the deformed heterotic 5-brane on a Taub-NUT
instanton, which preserves half of the original supersymmetry
\cite{clp}. It is a rather  pleasant surprise that wave equation of a
minimally-coupled, massless scalar propagating in this background can be
solved exactly and in closed-form in terms of hypergeometric functions. The
scalar dynamics are similar to that in the background of a two-charge black
hole in four dimensions. This enables us to study the deformation to a
nonsingular  spacetime from the point of view of perturbative field
dynamics.

While absorption probabilities can yield useful information on supergravity
backgrounds, additional motivation for studying absorption by $p$-branes 
is the conjectured duality between $d$-dimensional supergravity on certain
spacetimes with corresponding $d-1$-dimensional quantum field theories. 
While this conjecture has been most studied for the case of supergravity on
AdS spacetime \cite{malda}, such a duality may hold in a more general
context.

There are only a few known examples for which the wave equations of a 
scalar field in a $p$-brane background is exactly solvable. In the case of
the extremal D3-brane \cite{gubser,raams,muller,park}, as well as the
extremal D1-D5 intersection \cite{dyonic}, the wave equation can be cast
into the form of the modified Mathieu equation. The absorption probability
can then only be expressed in terms of power-series expansions of the
frequency. For the five-dimensional single-charge black hole and the
four-dimensional two-charge black hole the absorption probability can be
found in closed-form \cite{clpt,mass}.

This paper is organized as follows. In section 2 we present the wave
equation for a minimally-coupled, massless scalar propagating in the
background of a heterotic 5-brane on a Taub-NUT instanton. In section 3 we
find the conditions for non-zero absorption from the relative phase shift
between incident and reflected waves. In section 4 we calculate the
absorption probability exactly and in closed-form. In section 5 we analyze
the connection between absorption probability and radially-infalling
timelike geodesics. In section 6 we calculate the quasinormal modes
exactly.

\section{Scalar perturbations of heterotic 5-brane on Taub-NUT instanton}

We begin by briefly reviewing the deformed Heterotic 5-brane solution
found in \cite{clp}. The bosonic sector of the ten-dimensional heterotic
supergravity consists of the metric, a dilaton, a two-form potential
$A_{(2)}$ and the Yang-Mills fields of $E_8 \times E_8$ or $SO(32)$. The
corresponding Lagrangian is
\begin{equation}
L_{\textrm{het}}=R \ast \textbf{1}-\frac{1}{2}\ast d\phi
\wedge d\phi-\frac{1}{2}\textrm{e}^{-\phi}\ast F_{(3)} \wedge
F_{(3)}-\frac{1}{2}\textrm{e}^{-\frac{1}{2}\phi}\ast F_{(2)}^i
\wedge F_{(2)}^i,
\end{equation}
where
$$
F_{(3)} = dA_{(2)}+\frac{1}{2}A_{(1)}^i \wedge
dA_{(1)}^i+\frac{1}{6}f_{ijk}A_{(1)}^i \wedge A_{(1)}^j \wedge
A_{(1)}^k,
$$
\begin{equation}
F_{(2)}^i = dA_{(1)}^i+\frac{1}{2}f_{jk}^i A_{(1)}^j \wedge
A_{(1)}^k.
\end{equation}
The heterotic 5-brane deformed by an Abelian $U(1)$ field is
$$
ds_{10}^2=H^{-1/4}dx_{\mu}dx^{\nu}\eta_{\mu \nu}+H^{3/4}ds_4^2,
$$
\begin{equation}
\textrm{e}^{-\phi}\ast F_{3}=d^6 x\wedge dH^{-1},\ \ \
\phi=\frac{1}{2}\log H,\ \ \ F_{(2)}=mL_{(2)},
\end{equation}
where $L_{(2)}$ is a self-dual harmonic 2-form in the Ricci-flat
K\"{a}hler transverse metric $ds_4^2$. For the case of a Taub-NUT
instanton,
\begin{equation}
ds_4^2=\big( \frac{r+a}{r-a}\big) dr^2+4a^2\big(
\frac{r-a}{r+a}\big) d\psi^2+(r^2-a^2)(d\theta^2+\sin^2 \theta
d\phi^2),
\end{equation}
where the radial coordinate has the range $a \le r \le \infty$. The
normalisable solution for $L_{(2)}$ is given by
\begin{equation}
L_{(2)}^2= \frac{4}{(r+a)^4},
\end{equation}
for which half the original supersymmetry is preserved. The
Bianchi identity for $F_{(3)}$ is given by
\begin{equation}
dF_{(3)}=\frac{1}{2}F_{(2)}^i \wedge F_{(2)}^i.
\end{equation}
Solving for the function $H$, we obtain
\begin{equation}
H=1-\frac{4ab+m^2}{4a(r-a)}+\frac{m^2}{4a(r+a)}, \label{H}
\end{equation}
where $b$ is an integration constant. For the case $b=-m^2/(4a)$,
the function $H$ becomes non-singular in the entire radial
coordinate range \cite{clp}.

The equation of motion for a minimally-coupled scalar field is
\begin{equation}
\frac{1}{\sqrt{-g}}\partial_{\mu}\sqrt{-g}g^{\mu \nu}\partial_\nu \Phi=0.
\end{equation}
Taking
\begin{equation}
\Phi(t,r,\theta,\phi,\psi)=\phi(r)Y_{\ell,m}(\theta,\phi){\rm 
e}^{in\psi}{\rm e}^{-i\omega t},
\end{equation}
we find the radial equation to be
\begin{equation}
\frac{1}{r^2-a^2} \partial_r(r-a)^2\partial_r\phi+\big[
\omega^2H-\frac{r+a}{r-a}\frac{n^2}{4a^2}-\frac{\ell(\ell+1)}{r^2-a^2}\big]
\phi=0. \label{wave1}
\end{equation}
Making the wave function and coordinate transformations
$\phi(r)=(r-a)^{-1}\psi(r)$ and $z=r-a$, together with (\ref{H}), the
radial wave equation (\ref{wave1}) becomes
\bea
\partial_z^2\psi+\Big[
(\omega^2-\frac{n^2}{4a^2})+\Big(\omega^2(2a+\frac{m^2}{4a}-\frac
{4ab+m^2}{4a})-\frac{n^2}{a}\Big)\frac{1}{z}-\\
\Big(\frac{\omega^2}{2}(4ab+m^2)
+n^2+\ell(\ell+1)\Big)\frac{1}{z^2}\Big]\psi=0. \label{wave2} 
\eea 
This equation can be solved exactly in terms of hypergeometric
functions. Note that wave-like solutions exist only for $\omega > n/2a$.

\section{Phase shifts and absorption}

Before solving for the absorption probability, it is instructive to examine
the relative phase shift between incident and reflected waves, which is
obtained directly from the exact wave function solution. If the phase shift
has an imaginary component, then there is a nonzero absorption probability
\cite{bl}.

\subsection{Non-singular geometry}

In the case of a non-singular background for which $b=-m^2/(4a)$, the wave
equation (\ref{wave2}) is
\be
\partial_z^2\psi+\Big[\big( \omega^2-\frac{n^2}{4a^2}\big) 
+\big( (2a-b)\omega
^2-\frac{n^2}{a}\big)\frac{1}{z}-\frac{\big( n^2+\ell(\ell+1)\big) }{z^2}
\Big]\psi=0.
\ee
This wave equation is of the same form as that for a four-dimensional
single-charge black hole. It is also formally equivalent to the
Schr\"{o}dinger equation for a particle of energy $E=\omega^2-n^2/(4a^2)$ in
an attractive Coulomb potential of charge $Q=(2a-b)\omega^2-n^2/a$ with an 
effective angular momentum given through $L(L+1)=\ell(\ell+1)+n^2$. The exact
solution is
\be
\psi=N (\sqrt{E}z)^{L+1}{\rm e}^{-i\sqrt{E}z} M(\frac{iQ}{2\sqrt{E}}+L+1,
2L+2,2i\sqrt{E}z),
\ee
where $M$ is Kummer's regular confluent hypergeometric function and the
factor $N$ ensures that the solution is Dirac normalized. The asymptotic form
for large $z$ is
\be
\psi \sim \sin \Big( \sqrt{E}z+\frac{Q}{2\sqrt{E}}\log \sqrt{E}
z-\frac{L \pi}{2}+\delta \Big),
\ee
where
\be
\delta={\rm arg}\ \Gamma (L+1-i\frac{Q}{2\sqrt{E}}). \label{delta}
\ee
The relative phase shift between incident and reflected waves is
\be
2\Delta=-L\pi+2\delta.
\ee
$\Delta$ is real, which implies that the incoming and outgoing flux at
infinity are equal. Thus, the absorption probability is zero at all
frequencies, which has been well-established for Coulomb potentials, even
those that are attractive \cite{bl}.

\subsection{Singular geometry}

The above analysis can be repeated for arbitrary $b$, which include the case
of singular geometry. The wave equation remains formally identical to that of
the Schr\"{o}dinger equation for a particle of energy $E$ in an attractive
Coulomb potential. The effective charge is given by
\be
Q=\Big( 2a+\frac{m^2}{4a}-\frac{4ab+m^2}{4a}\Big) \omega^2-n^2/a,
\ee
and the effective angular momentum is given through
\be
L(L+1)=\ell(\ell+1)+n^2+(4ab+m^2)\frac{\omega^2}{2}.
\ee
L develops an imaginary component for
\be
\frac{\epsilon \omega^2}{2}>(\ell+1/2)^2+n^2,
\ee
where $b=-m^2/(4a)-\epsilon$. Nevertheless, the solution and calculation of
the phase shift remains valid for complex $L$ and we find that
\be
2\Delta=-L\pi+2\delta,
\ee
where $\delta$ is given by (\ref{delta}). When $L$ has a nonzero imaginary
component, the phase shift is complex and the incoming flux is not equal to
the outgoing flux. This indicates that the absorption probability is nonzero
for $\epsilon \omega^2/2>(\ell+1/2)^2+n^2$. That is, for the singular
geometry for which $\epsilon >0$, there is nonzero absorption above a certain
threshold frequency for the incident scalar wave. Note that it is the
waves of higher frequency that penetrate through the potential barrier
surrounding the singularity. Thus, it is as though nature cloaks the
singularity from the outside world via absorption.

\section{Closed-form absorption probability}

In general, the wave equation of a massless scalar in the background of a
p-brane is independent of the world-volume dimension $d$. This implies
that the wave equation is invariant under double-dimensional reduction of
the corresponding $p$-brane. For example, a scalar in the background of
an M2-M5 intersection in 11 dimensions and a six-dimensional dyonic string
shares the identical wave equation as for the D1-D5 intersection. However,
since a two-charge black hole in four dimensions can be obtained from the
D1-D5 intersection only via both vertical as well as diagonal reduction
steps, the corresponding scalar wave equation changes. The corresponding
wave equation is exactly solvable in terms of hypergeometric functions, and 
the absorption probability can be expressed in closed form. Another
interesting limit is when the D1-brane disappears, in which case the wave
equation in the D5-brane background is exactly solvable in terms of Bessel
functions. This is a second case for which the absorption probability can
be obtained in closed form. Dimensionally reducing the D5-brane down to a
five-dimensional single-charge black hole does not change the wave equation
\cite{clpt}.

The wave equation for a minimally-coupled, massless scalar
propagating in the background of a heterotic 5-brane on a Taub-NUT
instanton is identical to such a wave equation in the background of a
four-dimensional two-charge black hole. To make this apparent, we make the
wave function transformation $\psi=z\phi$ and the coordinate
transformation $\rho =\sqrt{\omega^2-n^2/(4a^2) } z$, the latter of which
is valid for $\omega > n/(2a)$. The wave equation can be expressed as
\be
\partial_{\rho}^2\phi+\frac{2}{\rho}\partial_{\rho}\phi+\Big[ \big(
1+\frac{\lambda_+}{\rho}\big) \big(
1+\frac{\lambda_-}{\rho}\big) -\frac{L(L+1)}{\rho^2}\Big] \phi=0. 
\ee
This is of the form of a minimally-coupled scalar field of effective
angular momentum given by $L(L+1)=\ell(\ell+1)+n^2$ propagating in the
background of a four-dimensional black hole with two charges given by
\be
\lambda_{\pm}=\frac{(2a-b)\omega^2-n^2/a}{\sqrt{4\omega^2-n^2/(a^2)}} \pm
\sqrt{\frac{\Big( (2a-b)\omega^2-n^2/a
\Big)^2}{4\omega^2-n^2/(a^2)}+(4ab+m^2)\frac{\omega^2}{2}}.
\ee
From above, we can see that the scalar perturbative dynamics in 
the background of the non-singular geometry $\epsilon=0$ are equivalent to
that in the background of a four-dimensional single-charge black hole, for
which it has been shown that the absorption probability is zero
\cite{clpt}. 

The wave function can be solved exactly with hypergeometric 
functions yielding 
\be
\phi=\rho^{(iq-1)/2}{\rm e}^{-i\rho} \Big( \alpha
U(\frac{1}{2}+ip+\frac{i}{2}q,
1+iq,2i\rho)+\beta M(\frac{1}{2}+ip+\frac{i}{2}q,1+iq,2i\rho) \Big),
\label{solution} 
\ee
where $U$ and $M$ are Kummer's irregular and regular confluent
hypergeometric
functions, respectively. We have defined $p \equiv \frac{1}{2}(\lambda_+
+\lambda_-)$ and $q \equiv \sqrt{4\lambda_+ \lambda_- -(2L+1)^2}$ as in
\cite{clpt}. From the asymptotic behavior of this solution, it can be seen
that in order to have wave-like behavior near the brane as well as
asymptotically far, we require that 
\be
2a\epsilon \omega^2 > (\ell+1/2)^2+n^2 \label{condition1}
\ee
and
\be
\omega > n/(2a), \label{condition2}
\ee
respectively. 

One can calculate the closed-form absorption probability to be
\be
P=\frac{1-{\rm e}^{-2\pi \sqrt{8a\epsilon \omega^2-1}}}{1+{\rm e}^{-\pi
\big( [2a+m^2/(4a)+\epsilon]\omega +\sqrt{8a\epsilon
\omega^2-1}\big) }},
\ee
if the conditions (\ref{condition1}) and (\ref{condition2}) are
satisfied; otherwise, $P=0$. This calculation has already been done
explicitly in \cite{clpt} for the analogous case of the two-charge black
hole. 

\section{Geodesics and absorption}

In the case of the extremal $D=5$ single-charge and $D=4$ two-charge
black holes, it has already been shown that the vanishing of the
absorption probabilities below certain threshold frequencies is related
to the behavior of geodesics in these backgrounds \cite{clpt}. That is,
if a timelike or null geodesic reaches $r=a$ in finite coordinate
time, $P=0$ for all frequencies. If it takes a logarithmically-divergent
coordinate time, $P>0$ for $\omega>\omega_0$, where $\omega_0$ is some
finite frequency. If it takes a power-law-divergent coordinate time,
$P>0$ for $\omega>0$. In this paper, we discuss examples of the first
two cases.

Consider radially-infalling timelike geodesics in the background of a
heterotic 5-brane on a Taub-NUT instanton. These are described by the
Lagrangian
\be
L=\frac{1}{2}g_{\mu \nu}\frac{dx^{\mu}}{d\tau}\frac{dx^{\nu}}{d\tau}=
-H^{-1/4}\Big( \frac{dt}{d\tau}\Big)^2+\Big(
\frac{r+a}{r-a}\Big) H^{3/4} \Big( \frac{dr}{d\tau}\Big)^2. 
\ee
The equation of motion for $t$ yields
\be
\frac{dt}{d\tau}=EH^{1/4}, \label{t}
\ee
where $E$ is a constant of integration. $L=-1$ for a timelike geodesic,
which yields 
\be
\Big( \frac{dr}{d\tau}\Big)^2=E^2\Big( \frac{r-a}{r+a}\Big) H^{-1/2}
-\Big( \frac{r-a}{r+a}\Big) H^{-3/4}.
\ee
For the non-singular case $\epsilon=0$, as we approach $r=a$, $H
\sim$ constant. Thus,
\be
\frac{dr}{dt}=\frac{dr}{d\tau}\frac{d\tau}{dt}\sim -(r-a)^{1/2}.
\ee
We find that timelike or null geodesics reach $r=a$ in a finite
coordinate time.

For the singular case of nonvanishing $\epsilon$, near $r=a$ we
have $H \sim (r-a)^{-1}$. Thus,
\be
\frac{dr}{dt}\sim -(r-a).
\ee
We find that the time taken for timelike or null geodesics to reach $r=a$ 
is logarithmically divergent. Thus, our results are in agreement
with the correpondence between the behavior of infalling geodesics and
absorption. This correspondence may be a useful tool to study the
absorption of fields in singular/non-singular backgrounds for which we are 
unable to calculate absorption probabilities analytically.

\section{Quasi-normal modes}

Note that the wave equation (\ref{wave2}) is of the form
\be
\partial_z^2 \psi+\Big(
A+\frac{B}{z}-\frac{C}{z^2}\Big) \psi=0. \label{form}
\ee
The conditions for wave-like solutions (purely real $\omega$),
(\ref{condition1}) and (\ref{condition2}), can be expressed as $A > 0$
and $C < -1/4$. These conditions impose the constraint that
\be
B > \frac{1}{2a}\Big( \frac{m^2 n^2}{8a^2}+(\ell+1/2)^2\Big).
\ee
However, one way that the wave equation (\ref{form}) differs from
those in standard $p$-brane backgrounds is that $B$ (analogous to the
$p$-brane charge) is not necessarily positive. Thus, by restricting
ourselves to solutions of purely real $\omega$ we may be missing some
interesting dynamical features. In this vein, we compute exactly the 
quasinormal modes \cite{birm,card1}.

For large $\rho$, the wave function solution (\ref{solution}) is
\be
\phi \sim \rho^{-ip-1}(-A{\rm e}^{-i\rho}+{\rm e}^{i\rho}),
\ee
with $A$ given by \cite{clpt}
\be
A=\frac{\Gamma (\frac{1}{2}+ip+\frac{i}{2}q)({\rm e}^{\pi q}+{\rm e}^{-2\pi
p})}{2i\Gamma(\frac{1}{2}-ip+\frac{i}{2}q)(2\rho)^{2ip}{\rm
e}^{-\pi p} \cosh \pi (p-\frac{1}{2}q)}.
\ee

A quasinormal mode is a free oscillation of the brane itself, with no
incoming radiation driving it. Thus, such modes are defined as solutions 
which are purely ingoing at the horizon and purely outgoing at infinity
\footnote{For asymptotically AdS spacetimes, quasinormal modes must vanish
at infinity}. Thus, for quasinormal modes, we require that $A=0$. This
imposes the restriction
\be
\frac{1}{2}-ip+\frac{i}{2}q=-N, \label{quasi}
\ee
where $N=0,1,2,..$ From (\ref{quasi}), we find the exact quasinormal mode 
frequencies. For simplicity, we consider the case $n=0$: 
\bea
\omega_{\pm}=\frac{-i(2N+1)(2a-b)}{\Delta}\\ \nonumber 
\pm \frac{\sqrt{-(2N+1)^2 (2a-b)^2+ [(2N+1)^2-
(2\ell+1)^2]\Delta}}{\Delta}, \label{freq}
\eea
where $\Delta \equiv (2a+b)^2+2m^2>0$. 

In order for the geometry to remain stable in the presence of quasinormal
modes, we must have ${\rm Im}\ \omega <0$. Note that ${\rm Im}\ \omega$
describes the decay of the scalar perturbation. This offers a prediction of
the timescale for return to equilibrium of the dual quantum field theory,
if one exists \cite{birm}.

From the above equation for $\omega_{\pm}$, we see that the geometry is
stable against such perturbations if $2a-b>0$, which implies that the
effective charge $B>0$ in (\ref{form}) (we consider modes such that the
term in the square root of (\ref{freq}) is positive, so that ${\rm Re}\
\omega>0$). Note that, for the non-singular geometry where $b=-m^2/(4a)$, 
$\omega$ becomes completely imaginary.

\section*{Acknowledgments}

We are grateful to Mirjam Cveti\v{c} and Marc Henneaux for useful
discussions.

\end{document}